\documentclass[10pt,leqno]{amsart}
\usepackage{graphicx}
\usepackage{indentfirst,csquotes}

\usepackage{amssymb,amsthm,amsmath}
\usepackage{xcolor,paralist,hyperref,fancyhdr,etoolbox}
\usepackage{amsmath,amsthm}
\usepackage{xr}
\usepackage{xcite}
\usepackage{amsfonts}
\usepackage{enumerate}
\usepackage{natbib}
\usepackage{cases}
\usepackage{booktabs}
\usepackage{float}

\numberwithin{equation}{section}
\newtheorem{thm}{Theorem}

\newtheorem{prop}{Proposition}
\theoremstyle{definition}

\theoremstyle{remark}
\newtheorem{rem}{Remark}
\newtheorem{ass}{\textbf{Assumption}}

\newtheorem{example}{Example}
\allowdisplaybreaks[2]

\newcommand{\Em}{\mathbb{E}}
\newcommand{\Pm}{\mathbb{P}}

\begin{document}
\title[Bivariate Heteroscedastic Extremes]{Bootstrap-based Inference for Bivariate Heteroscedastic Extremes with a Changing Tail Copula} 
\author{Yifan Hu$^{\ast}$ }
\author{Yanxi Hou$^{ \ast,\dagger}$}
\footnotetext[1]{ 
 School of Data Science, Fudan University, Shanghai, China, 200433  
}
\footnotetext[2]{ Corresponding author, \href{Email:email-id.com}{yxhou@fudan.edu.cn} }
\date{\today}

\begin{abstract}
      This paper introduces a copula-based model for independent but non-identically distributed data with heteroscedastic extremes marginal and changing tail dependence structures. We establish a unified framework for inference by proving the weak convergence of the bivariate sequential tail empirical process (B-STEP) and its empirical bootstrap counterpart. We derive the asymptotic properties of several estimators on the tail, including the quasi-tail copula, integrated scedasis function, and Hill estimator, treating them as functionals of B-STEP. This process-centric approach enables the development of bootstrap-based methods and ensures the theoretical validity of the derived statistics. As an application of our inference method, we propose bootstrap-based tests for the equivalence of extreme value indices, the equivalence of scedasis functions, and non-changing tail dependence when marginal scedasis functions are identical. Our simulations validate the robustness and efficiency of the bootstrap-based tests. 
\end{abstract} 

\maketitle
\noindent%


\section{Introduction}\label{sec:introduction}

Recent advancements in extreme value theory (EVT) have broadened its scope from independent and identically distributed (IID) data to independent but non-identically distributed (IND) data. Within this IND framework, researchers have delved into univariate heterogeneous extremes. For instance, \citet{Zhou2016} introduced a scedasis function to capture the variability in tail probabilities. Building on this foundation, \citet{einmahl2022extreme} and \citet{Einmahl2023} further extended the framework to accommodate more diverse forms of heterogeneity in extreme values. Concurrently, other studies have addressed challenges such as trend detection \citep{Mefleh2020} and time-series dependencies \citep{Jennessen2024}. These studies have demonstrated the promising potential for applying heterogeneous extremes in complex modeling and real-world applications.

Establishing EVT under the bivariate IND framework is challenging due to two sources of heterogeneity: marginal distributions and dependence structures. On the one hand, \citet{einmahl2024tail} proposes a framework for marginal heterogeneity but assumes a common tail copula structure across samples, with heteroscedastic extremes for each dimension. On the other hand, research into changing tail copulas is a recent development, with \citet{drees2023statistical} addressing the specific problem of testing for a non-changing copula structure.
Whereas existing IND frameworks focus on either marginal heterogeneity or non-stationary dependence, our work introduces a copula-based model that integrates both heterogeneity in marginal and dependence structures. This joint modeling is critical for applications where marginal and dependence trends evolve simultaneously, such as climate extremes (e.g., rising temperatures combined with shifting storm dependencies) or financial risk (e.g., evolving tail risks alongside contagion effects). Additionally, this framework is important for the joint inference of marginal statistics, such as the inference for the equivalence of the scedasis functions and the two-sample tests for the extreme value index (EVI).

Moreover, the inference of our proposed model is challenging, as the asymptotic variance structures involve unknown functions introduced by marginal and dependence heterogeneity on the tail region. This phenomenon has also been identified in prior research. For example, \citet{drees2023statistical} states that their integrated angular measure for testing tail dependence changes is influenced by marginal distributions when the same extreme observations are used for both marginal and dependence estimation. Additionally, we find that the Kolmogorov-Smirnov (KS) statistic for testing the equality of scedasis functions \citep{einmahl2024tail} cannot be adjusted to a standard Brownian Bridge when the copula structure varies across samples. Consequently, this challenge complicates theoretical research on bivariate heteroscedastic extremes, an area that is still largely unexplored in the literature.

To address the intricate inference problem arising from the complex asymptotic variance structure in the framework of heteroscedastic extremes, the bootstrap method is widely employed and holds substantial theoretical value. Recent advancements have introduced bootstrap approaches for EVT under IID and serially dependent assumptions, targeting estimators such as probability weighted moment (PWM) estimators \citep{haan2024}, Hill estimators \citep{Jentsch2021}, and tail copulas \citep{Bucher2013}. For example, \citet{Bucher2013} apply transformations to bootstrap empirical processes to facilitate inference for tail copulas in IID settings. Similarly, within the IND framework where estimators are linked to the \textit{sequential tail empirical process} (STEP) \citep{Zhou2016}, developing a bootstrap STEP is crucial for constructing effective bootstrap methods, which motivates this paper.

 To study the joint inference on marginal and dependency heterogeneity, we model the survival distribution function $S_{n, i}$
 of the bivariate sample $(X_{i}^{(n)}, Y_{i}^{(n)})$ for $i = 1,\ldots, n$ by Sklar's theorem, and assume a (survival) copula $C_{n, i}$ satisfying for the two marginal distributions $F_{n, i}^{(j)}$, $j=1,2$, such that
\begin{equation}\label{eq:sklar2}
 S_{n,i}(x,y) = C_{n,i}(1-F_{n,i}^{(1)}(x),1-F_{n,i}^{(2)}(y)), \quad (x,y)\in\mathbb R^2.
\end{equation}
 Moreover, the two marginal distributions
$\{F_{n,i}^{(1)}\}_{i=1}^n$ and $\{F_{n,i}^{(2)}\}_{i=1}^n$ follows univariate
 heteroscedastic extremes. More specifically, there exists a heavy-tailed distribution function $G_j$ and a {\it scedasis function} $c_j$
such that for all $i = 1,\ldots, n$ and $n \in \mathbb{N}$, 
\begin{equation}\label{eq:heter}
 \lim_{t\to \infty} \frac{1 - F^{(j)}_{n,i}(t)}{1 - G_j(t)} = c_j\left(\frac{i}{n}\right),\quad j=1,2,
\end{equation}
where $c_j$ is positive and continuous subject to the constraint $\int_{0}^{1} c_j(s) ds = 1$ for $j = 1,2$.
$C_j(z)=\int_0^zc_j(s)ds$ is called the {\it integrated scedasis function}.
Without an explicit statement in our paper, the two scedasis functions are not assumed to be identical. 
For the changing copula structure, we assume a positive function 
$R$ on $\mathbb{R}^3$ 
satisfying for all $i = 1,\ldots, n$ and $n \in \mathbb{N}$, 
\begin{equation}\label{eq:tcind2}
 \lim_{t\to\infty} tC_{n,i}\left(\frac{x}{t}, \frac{y}{t}\right) = R\left(x, y, \frac{i}{n}\right), \quad (x,y) \in \mathbb{R}^2.
\end{equation}
Here, $R(x,y,z)$ is a tail copula given any $z$ and captures the changing tail dependence across samples, which generalizes the common tail copula in \cite{einmahl2024tail}.
We refer to this model as \textit{Bivariate Heteroscedastic Extremes}, which extends the concept of heteroscedastic extremes to incorporate a changing tail dependence structure. More detailed assumptions and discussions about this model are provided in Section \ref{sec:BHE}.

Based on the proposed model, this paper establishes a unified framework for inferring marginal estimators and tail dependence structures by proving the weak convergence of the \textit{bivariate sequential tail empirical process} (B-STEP). Additionally, we develop a bootstrap B-STEP that jointly resamples margins and copulas, accommodating both sources of heterogeneity. Using the convergence results of B-STEP, we derive the asymptotic statistical properties of several important statistics that are functionals of B-STEP. Specifically, we examine three key statistics: the quasi-tail copula, the integrated scedasis function, and the Hill estimator. For each, we establish the corresponding asymptotic theorems and the properties of their bootstrap counterparts. Moreover, we employ the bootstrap method to address three testing problems: the equivalence of the EVIs on different margins, the equivalence of two scedasis functions on different margins, and the constancy of the copula structure across samples when the marginal scedasis functions are identical. We establish the asymptotic theory for these tests, ensuring that the rejection probabilities converge to the specified significance level as the sample size and bootstrap replications diverge to infinity.

The rest of this paper is organized as follows. 
In Section \ref{sec:BHE}, we provide details of the assumptions of Bivariate Heteroscedastic Extremes.
In Section \ref{sec:tailst}, we introduce the B-STEP and its bootstrap process and develop estimators as functionals of the B-STEP.
In Section \ref{sec:inference}, we establish the asymptotic 
convergence results and then provide joint asymptotic properties of the estimators. 
In Section \ref{sec:Test}, we examine three hypothesis tests and provide the simulation results. 
Proofs are included in the Supplementary Material.

\section{Bivariate Heteroscedastic Extremes}
\label{sec:BHE}
In this section, we introduce the assumptions for the bivariate heteroscedastic extremes model,
\begin{equation}\label{mod: bind}
 \left\{\begin{array}{l}
 X_i^{(n)}\overset{}{\sim} F_{n,i}^{(1)},\quad
 Y_i^{(n)}\overset{}{\sim} F_{n,i}^{(2)}, 
 \text{where } F_{n,i}^{(j)} \text{ satisfies \eqref{eq:heter} for } j=1,2, \\
 \left(1-F_{n,i}^{(1)}(X_i^{(n)}), 1-F_{n,i}^{(2)}(Y_i^{(n)})\right)
 \overset{}{\sim} C_{n,i}, 
 \text{with } C_{n,i} \text{ satisfying \eqref{eq:tcind2}}.
    \end{array}\right.
\end{equation}

Assumption \ref{ass:marginal} characterizes the tail behavior of the marginal distributions \( F_{n, i}^{(j)} \). These conditions are consistent with those established in \citet{Zhou2016} for univariate distributions.
\begin{ass}[Marginal Heteroscedastic Extremes]\label{ass:marginal}
 For $j = 1,2$, the scedasis function $c_j(s)$ is positive, continuous, and bounded away from $0$ on $[0,1]$, satisfying $C_j(1) = 1$. Moreover,  there exist a positive, eventually decreasing functiona $A_j$ with $\lim _{t \rightarrow \infty} A_j(t)=0$, and a distribution function $G_j$, such that as $t \rightarrow \infty$,
              $$
 \sup _{n \in \mathbb{N}}
 \max _{1 \le i \le n}
 \left|
 \frac{1-F^{(j)}_{n,i}(t)}{1-G_j(t)}-
 c_j\left(\frac{i}{n}\right)
 \right|
 =O\left[A_j\left\{\frac{1}{1-G_j(t)}\right\}\right] .
              $$
    For the reference function $G_j$,  there exist some $\gamma_j >0,\beta_j <0$, an eventually positive or negative function $B_j$, such that
              $$
                  \begin{aligned}
                       & \lim _{t \rightarrow \infty}
 \frac{1}{B_j\left(1 /(1 - {G}_j(t))\right)}
 \left(\frac{1 - {G}_j(t x)}{1-{G}_j(t)}
 -x^{-1 / \gamma_j}\right)
 =x^{-1 / \gamma_j}
 \frac{x^{\beta_j / \gamma_j}-1}{\gamma_j \beta_j},
 \quad x>0 \,.
                  \end{aligned}
              $$
    
\end{ass}

Assumption \ref{ass:cop} specifies the convergence rate of \eqref{eq:tcind2}. This condition extends the assumptions in \cite{einmahl2024tail} to accommodate the modeling of a changing copula structure. Specifically, the tail copula $R$ is required to be non-degenerate at each $z \in [0,1]$. For every fixed $z$, the conditions imposed on the derivatives of $R$ are analogous to those discussed in \citet{Bucher2013}. 

\begin{ass}[Tail Dependence Structure]\label{ass:cop}  
    $R(x,y,z)$ is continuous on $\mathbb{R}^3$ satisfying $R(1,1,z) > 0$ for all $z \in [0,1]$. The partial derivatives of $R$ exist and satisfy that
\begin{align*}
    {\partial R}(x,y,z)/{\partial x}& \,\text{ is continuous on } 0 < x < \infty, 0 \le y < \infty; \\
    {\partial R}(x,y,z)/{\partial y}& \,\text{ is continuous on } 0 < y < \infty, 0 \le x < \infty.
\end{align*}
 Moreover, it holds for an $\alpha > 0$ and a constant $T \ge 1$, that as $t \to \infty$,
    \begin{equation}\label{eq: tailstr}
 \sup_{n \in \mathbb{N}} \sup_{\substack{0 <x,y \le T \\ i = 1, \ldots, n}} \left| tC_{n,i}\left( \frac{x}{t}, \frac{y}{t}\right)  - R\left(x,y, \frac{i}{n}\right) \right| = O\left( t^{-\alpha} \right).
    \end{equation}
\end{ass}

Assumption \ref{ass: smoothing} generalizes the smoothing conditions presented in \cite{Zhou2016} to the bivariate context. Furthermore, to demonstrate the practicality of these assumptions, we provide two examples that meet both Assumptions \ref{ass:cop} and \ref{ass: smoothing}.

\begin{ass}[Smoothing Conditions] \label{ass: smoothing}
 For the scedasis functions and tail copula functions, it holds for $j = 1,2$, that 
    \begin{align*}
 \lim _{n \rightarrow \infty}  \sup _{|u-v| \le 1 / n} \sqrt{k} | c_j(u)-c_j(v)| & =0, \\
 \lim_{n\to\infty} \sup_{\substack{|u-v| \le 1/n,\, y \in [0,T].}} \sqrt{k}\left| R(1,y,u) -R(1,y,v) \right| & = 0, \\
 \lim_{n\to\infty} \sup_{\substack{|u-v| \le 1/n,\, x \in [0,T].}} \sqrt{k}\left| R(x, 1,u) -R(x,1, v) \right| & = 0.
    \end{align*}
\end{ass}

\begin{example}[Mixture Copula]\label{exm: Mixture} Denote \(C_1(u,v) = (u^{-1} + v^{-1} - 1)^{-1}\) as a Clayton copula, and  \(C_2(u,v) = u + v - 1 + {(1-u)(1-v)}/({1 - uv})\) as a Ali-Mikhail-Haq copula.
 A changing (survival) copula is given by
 \[
 C_{n,i}(u,v) := p(i/n) C_1(u,v) + (1 - p(i/n)) C_2(u,v), \quad (u,v) \in [0,1]^2,
\]
where  \(0 < p(z) \le 1\) is a function satisfying $
 \lim_{n \to \infty} \sqrt{k} \sup_{|u - v| \le 1/n} |p(u) - p(v)| = 0.
    $
The function $R$ is defined as \(
 R(x,y,z) = p(z) (x^{-1} + y^{-1})^{-1}.
 \) 
 
 We then prove that the mixture copula $C_{n,i}$ satisfies Assumption \ref{ass:cop}. For sufficiently large \(t > T\), we observe that
    \begin{align*}
        & \sup_n \sup_{\substack{0 \le x, y \le T \\ 1 \le i \le n}}
 \left| tC_{n,i}(xt^{-1}, yt^{-1}) - p(i/n) (x^{-1} + y^{-1})^{-1} \right| \\
 \le & \sup_n \sup_{\substack{0 \le x, y \le T \\ 1 \le i \le n}}
 (1 - p(i/n)) \left| x + y - t + \frac{t(1 - xt^{-1})(1 - yt^{-1})}{1 - xyt^{-2}} \right| \\
        & + \sup_n \sup_{\substack{0 \le x, y \le T \\ 1 \le i \le n}}
 \left| \frac{t p(i/n)}{t x^{-1} + t y^{-1} - 1} - \frac{p(i/n)}{x^{-1} + y^{-1}} \right| \\
 \le & \sup_{\substack{0 \le x, y \le T}}
 \left| \frac{xy^2 + x^2y - 2xy t}{xy - t^2} \right|
 + \left| \frac{1}{x^{-1} + y^{-1} - t(x^{-1} + y^{-1})^2} \right| \\
 \le & \frac{2T^2 t }{t^2 - T^2} + \frac{T^2/4}{t - T/2} \\
 = & O(1/t).
    \end{align*}

   Furthermore,  Assumption \ref{ass: smoothing} is satisfied if the function \(p\) meets the condition:
    \[
 \lim_{n \to \infty} \sqrt{k} \sup_{|u - v| \le 1/n} |p(u) - p(v)| = 0. \qedhere
    \]
 In this example, the mixture probability \(p(z)\) and the tail copula \((x,y) \mapsto (x^{-1} + y^{-1})^{-1}\) of the Clayton copula govern the tail dependence structure.
\end{example}

\begin{example}[Parameterized Copula Sequence]\label{exm:ParameterizedCopula}
 Consider the copula 
    \[
 C_{n,i}(u,v) = \left(u^{-\theta(i/n)} + v^{-\theta(i/n)} - 1\right)^{-1/\theta(i/n)}.
    \]
  If \(\theta\) is Lipschitz continuous, and if \(M_1 = \inf_{0 \le z \le 1} \theta(z) > 0\) and \(M_2 = \sup_{0 \le z \le 1} \theta(z) < \infty\), then the function $R$ is defined as \(R(x,y,z) = (x^{-\theta(z)} + y^{-\theta(z)})^{-1/\theta(z)}\). 

  We first verify that the parameterized copula sequence $C_{n,i}$ satisfies Assumption \ref{ass:cop}. For \(t > T\), the following uniform convergence holds:
  \begin{align*}
      & \sup_n \sup_{\substack{0 \le x, y \le T \\ 1 \le i \le n}}
\left| tC_{n,i}\left(\frac{x}{t}, \frac{y}{t}\right) 
- \left(x^{-\theta(i/n)} + y^{-\theta(i/n)}\right)^{-1/\theta(i/n)} \right| \\
= & \sup_{n \in \mathbb{N}} 
\sup_{\substack{0 < x, y \le T \\ i = 1, \ldots, n}} 
\left(x^{-\theta(i/n)} + y^{-\theta(i/n)}\right)^{-1/\theta(i/n)}
\frac{t - \left[t^{\theta(i/n)} - \frac{(xy)^{\theta(i/n)}}{x^{\theta(i/n)} + y^{\theta(i/n)}}\right]^{1/\theta(i/n)}}
{\left[t^{\theta(i/n)} - \frac{(xy)^{\theta(i/n)}}{x^{\theta(i/n)} + y^{\theta(i/n)}}\right]^{1/\theta(i/n)}} \\
\le & T \sup_{n \in \mathbb{N}} \sup_{\substack{i = 1, \ldots, n}} 
\left(1 - \left(1 - \frac{T^{\theta(i/n)}}{2 t^{\theta(i/n)}}\right)^{1/\theta(i/n)}\right) \\
\le & T \max\left(2^{1-1/M_2}, 1\right) \frac{T^{M_2}}{2 M_1 t^{M_1}}.
  \end{align*}
To verify that $C_{n,i}$ satisfies Assumption \ref{ass: smoothing},
we derive the derivative of $R$ with respect to \(\theta\), 
  \[
\left(y^{-\theta} + x^{-\theta}\right)^{-1/\theta}
\left(
\frac{\ln\left(y^{-\theta} + x^{-\theta}\right)}{\theta^2}
+ \frac{(\ln y) \cdot y^{-\theta} + (\ln x) \cdot x^{-\theta}}
{\theta \left(y^{-\theta} + x^{-\theta}\right)}
\right).
  \]
Due to the symmetry of the expression in \(x\) and \(y\), we set \(y = 1\) and obtain that
  \[
\left(1 + x^{-\theta}\right)^{-1/\theta}
\left(
\frac{\ln\left(1 + x^{-\theta}\right)}{\theta^2}
+ \frac{\ln(x) \cdot x^{-\theta}}
{\theta \left(1 + x^{-\theta}\right)}
\right),
  \]
which is bounded for \(x \in [0,1]\). 
Thus, by the Lipschitz continuity of $\theta(z)$, Assumption \ref{ass: smoothing} is satisfied.
 This example corresponds to the `G-linear' and `t-linear' models used in the simulation study of \cite{drees2023statistical}, where the copula structure is controlled by a changing parameter.
\end{example}

Assumption \ref{ass:intermediate} specifies the conditions for intermediate orders. It is important to note that we permit distinct intermediate order sequences $k_j$ for $j = 1,2$.

\begin{ass}[Intermediate Order]\label{ass:intermediate}
 The sequences $k$ and $k_j$ satisfy $k/n \to 0$, $k / k_j \to s_j > T^{-1}$, $\sqrt{k}A_j(n/(2Tk)) \to 0$, $\sqrt{k}B_j(n/k) \to 0$, and $\sqrt{k}(n/k)^{-\alpha} \to 0$ as $n\to\infty$ and for $j = 1,2$.
\end{ass}

\section{Estimators as Functions of Tail Processes}
\label{sec:tailst}

This section concentrates on the development of tail estimators and their associated processes under the assumptions in Section \ref{sec:BHE}. Theorems concerning the asymptotic properties of the estimators are postponed to the next section. We begin by introducing some necessary notations.
Let \(X_{k,n}\) denote the \(k\)-th smallest order statistic of \(X_{1}^{(n)}, \ldots, X_{n}^{(n)}\), and \(Y_{k,n}\) denote the \(k\)-th order statistic of \(Y_{1}^{(n)}, \ldots, Y_{n}^{(n)}\). 
We define 
\(
\lfloor x \rfloor := \max\{i \in \mathbb{Z} \mid i \le x\}\),
\(x \vee y = \max(x,y)\), and \(x \wedge y = \min(x,y)\).
The inverse of a non-decreasing function \(f\) is given by
\[f^{\leftarrow}(x) = \begin{cases}
\sup \left\{t \in \mathbb{R}_{+} \mid f(t)=0\right\}, & x=0, \\ 
\inf \left\{t \in \mathbb{R}_{+} \mid f(t) \geq x\right\}, & 0<x<\sup \operatorname{ran} f, \\ 
\inf \left\{t \in \mathbb{R}_{+} \mid f(t)=\sup \operatorname{ran} f\right\}, & x \geq \sup \operatorname{ran} f.\end{cases}
\] 
Specifically, the inverse function of \(1 / (1 - G_j)\) at \(t\) is denoted as \(U_j(t)\) for \(j = 1,2\). 
We introduce a weight function \(q(x,y)\) with a constant \(0 \le \eta < 1/2\) by
\begin{equation}\label{eq:q}
 q(x,y) = \begin{cases}
 (x \vee y)^\eta, & \text{if } (x, y) \in \mathbb{R}^2,   \\
 x^\eta , & \text{if } x \in \mathbb{R},\, y = \infty, \\
 y^\eta , & \text{if } y \in \mathbb{R},\, x = \infty.
  \end{cases}
\end{equation}
For simplicity, we denote \(x \mapsto q(x, \infty)\) and \(y \mapsto q(\infty, y)\) as \( q_1(x)\) and \( q_2(y)\), respectively. 
An important space in our analysis is defined as
\begin{align*}
 \mathbb{D}_{T}:= & \{(x,y,z) \mid 0 \le z \le 1, 0 \le x, y \le T\} \cup \{(x,y,z) \mid 0 \le z \le 1, x = \infty, 0 \le y \le T\} \\
& \cup \{(x,y,z) \mid 0 \le z \le 1, y = \infty, 0 \le x \le T\}.
\end{align*}
The space \(\ell^\infty(\mathbb{D}_{T})\) represents the set of bounded functions on \(\mathbb{D}_{T}\).

\subsection{Bivariate Sequential Tail Empirical Process}

In \citet{Zhou2016}, the STEP process serves as an important theoretical tool, which is defined as:
\begin{equation}\label{eq:stepuni}
 \frac{1}{k} \sum_{i=1}^{\lfloor nz \rfloor} \mathbf{1}\left\{X_i^{(n)} > U_1(n / (k x))\right\} - C_1(z) x.
\end{equation}
A nature extension is to analyze a process whose projection on the two margins is a STEP in the univariate context. Thus, we denote $\tilde{R}^\prime (x,y,z )$ as
\begin{equation}\label{eq: B-STEP}
 \tilde{R}^\prime (x,y,z ) := \frac{1}{k} \sum_{i=1}^{\lfloor nz \rfloor} \mathbf{1}\left\{X_i^{(n)} > U_1(n / (k x)), Y_i^{(n)} > U_2(n / (k y))\right\}.
\end{equation}
Suppose $1-F_{n,i}^{(j)}(t) \approx c_j(i/n)(1-F_j(t))$, the expectation of $\tilde{R}^\prime (x,y,z ) $ is, 
\begin{align*}
 \Em(\tilde{R}^\prime (x,y,z ))
  & \approx \frac{1}{n} \frac{n}{k} \sum_{i=1}^{\lfloor nz \rfloor} C_{n,i}\left(c_1\left(\frac{i}{n} \right) \frac{kx}{n}, c_2\left(\frac{i}{n} \right) \frac{ky}{n}\right) \\
  & \approx {R^\prime(x,y,z)} := \int_{0}^{z} R(c_1(t) x, c_2(t) y)\, dt  
 , \quad \text{as } n\to \infty.
\end{align*}
Given $z$, $R^\prime(x,y,z)$ is not necessarily a tail copula; therefore, we refer to $R^\prime(x,y,z)$ as {\it quasi-tail copula}. 
In risk modeling, ${R}^\prime(x, y, z)$ quantifies the tail probability that two assets simultaneously exceed intermediate thresholds, $U_1(n/(kx))$ and $U_2(n/(ky))$, during the time interval $[0, z]$. 
This characteristic makes ${R}^\prime(x, y, z)$ a dynamic index of extreme co-movements.

We then define the B-STEP with the weight function $q$ by
\begin{equation} \label{eq: step-general}
\mathbb{F}_n(x, y, z) = \frac{1}{q(x,y)} \left(\tilde{R}^\prime (x,y,z ) - R^\prime(x,y,z)\right).
\end{equation}

\begin{rem}
 Specifically, we calculate $R^\prime$ on $\mathbb{D}_{T}$ as:
  \[
 R^\prime(x,y,z) =
  \begin{cases}
    \int_{0}^{z} R(c_1(t)x, c_2(t)y, t) \, dt, & 0 \le x, y < \infty, 0\le z\le 1, \\
    \int_{0}^{z} R(c_1(t)x, \infty, t) \, dt = x \, C_1(z), & y = \infty, \, 0 < x < \infty, 0\le z\le 1, \\
    \int_{0}^{z} R(\infty, c_2(t)y, t) \, dt = y \, C_2(z), & x = \infty, \, 0 < y < \infty, 0\le z\le 1.
  \end{cases}
  \]
 Since for every fixed $z$, $R$ is a tail copula, the definition of $R^\prime$ aligns with $\int_{0}^{z} R(c_1(t)x, c_2(t)y, t) \, dt$ when extended to $\mathbb{D}_{T}$. 
 Moreover, following the notations in \cite{Bucher2013}, we denote the refined partial derivatives of $R^\prime$ to include the case when $x = 0$ or $y = 0$ by
  \[
 R^\prime_1 (x,y,z) =
  \begin{cases}
    0 , & \{(x,y,z) \in \mathbb{D}_{T} \mid x = 0 \text{ or } x= \infty\}, \\
 {\partial R^\prime}(x, y, z) / {\partial x}, & 
    \{(x,y,z) \in \mathbb{D}_{T} \mid 0 < x < \infty, 0\le y \le \infty\}, \\
  \end{cases}
  \]
  \[
 R^\prime_2 (x,y,z) =
  \begin{cases}
    0 , & \{(x,y,z) \in \mathbb{D}_{T} \mid y = 0 \text{ or } y= \infty\}, \\
 {\partial R^\prime}(x, y, z) / {\partial y}, & 
    \{(x,y,z) \in \mathbb{D}_{T} \mid 0 \le x \le \infty, 0 < y < \infty\}. \\
  \end{cases}
  \]
\end{rem}

We now formulate the bootstrap B-STEP. 
For a fixed index \(b\), let \(\{\xi_{bi}\}_{i=1}^n\) be an IID sequence of positive random variables, independent of \(\{(X_i^{(n)}, Y_i^{(n)})\}_{i=1}^n\). This sequence is replicated for \(b = 1, 2, \ldots, B\).
We define the bootstrap empirical distribution functions as
\[
\begin{aligned}
F_{b}^{(1)}(x) = \frac{1}{n} \sum_{i=1}^{n} \xi_{bi} \mathbf{1}\left(X_{i}^{(n)} \le x\right), \quad \text{and} \quad
F_{b}^{(2)}(y) = \frac{1}{n} \sum_{i=1}^{n} \xi_{bi} \mathbf{1}\left(Y_{i}^{(n)} \le y\right),
\end{aligned}
\]
and their corresponding generalized inverses \(U_{b}^{(j)} = \left(1 / (1 - F_{b}^{(j)})\right)^{\leftarrow}\) for \(j = 1, 2\). 
The bootstrap estimator of \eqref{eq: B-STEP} is given by
\[
\tilde{R}^{b\prime}(x, y, z) := \frac{1}{k} \sum_{i=1}^{\lfloor nz \rfloor} \xi_{bi} \mathbf{1}\left\{X_i^{(n)} > U_1(n / (k x)), Y_i^{(n)} > U_2(n / (k y))\right\}.
\]
The bootstrap B-STEP is then given by
\begin{equation} \label{eq: step-bootstrap}  
 \mathbb{F}^{b}_n(x, y, z) = \frac{1}{q(x,y)} \left(
 \tilde{R}^{b\prime}(x, y, z) - \tilde{R}^{\prime}(x, y, z)
 \right).
\end{equation}
The following assumption is for the bootstrap weights.

\begin{ass}[Bootstrap Weight]\label{ass:bootweight}
 The bootstrap weights \(\{\xi_{bi}\}_{i=1}^n\) satisfy that for the constant \(\eta > 0\) of $q$ in \eqref{eq:q}, we have
  \[
 \Em[\xi_{bi}] = 1, \quad \Em[(\xi_{bi} - 1)^2] = 1, \quad \text{and} \quad \Em[|1 - \xi_{bi}|^{1/\eta }] < \infty.
  \]
\end{ass}


\subsection{Estimator for the Quasi-Tail Copula.} 
As \(U_j\) is typically unknown, empirical quantiles are used to provide a data-driven approximation for tail thresholds. Specifically, the estimator for the quasi-tail copula is defined as
\begin{equation}\label{est:tailcop} 
 \hat{R}^\prime(x, y, z) = \frac{1}{k} \sum_{i=1}^{\lfloor nz \rfloor} \mathbf{1}\left(X_i^{(n)} > X_{n - \lfloor k x \rfloor, n}, Y_i^{(n)} > Y_{n - \lfloor k y \rfloor, n}\right).
\end{equation}  
The process \(\hat{R}^\prime\) can be derived from the process \(\tilde{R}^\prime\) by applying a functional $\Phi$ and the delta method to the process \(\mathbb{F}_n\), such that
$
\Phi({\tilde{R}^\prime}/{q}) - \Phi\left({R^\prime}/{q}\right) = \hat{R}^\prime - R^\prime$. For a function \(\theta\) on $\mathbb{D}_{T}$ satisfying \(\theta(0+, \infty, 1) = 0\), \(\theta(\infty, 0+, 1) = 0\), and that $\theta(x, \infty, 1)$ and $\theta(\infty, y, 1)$ are non-decreasing on $[0,T]$, $\Phi$ is given by
\begin{equation}\label{eq: phi} 
  \Phi(\theta)(x, y, z) = 
  \begin{cases} 
 (q \cdot \theta)\left((q \cdot \theta)^{\leftarrow}(x, \infty, 1), (q \cdot \theta)^{\leftarrow}(\infty, y, 1), z\right) & \text{if } x, y \neq \infty, \\ 
 (q \cdot \theta)\left((q \cdot \theta)^{\leftarrow}(x, \infty, 1), \infty, z\right) & \text{if } y = \infty, \\ 
 (q \cdot \theta)\left(\infty, (q \cdot \theta)^{\leftarrow}(\infty, y, 1), z\right) & \text{if } x = \infty. 
  \end{cases} 
\end{equation} 
Notice for $R'$, it holds that 
\(
R^{\prime \leftarrow}(x, \infty, 1) = x\), and \( R^{\prime \leftarrow}(\infty, y, 1) = y. 
\) 
Moreover, the inverse of $\tilde{R}^{\prime}(x, \infty, 1) $ and $\tilde{R}^{\prime}( \infty, y, 1) $ satisfy that
\[
\tilde{R}^{\prime \leftarrow}(x, \infty, 1) = \frac{n}{k} \{1 - G_1(X_{n, n - \lfloor k x \rfloor})\}, 
\quad 
\tilde{R}^{\prime \leftarrow}(\infty, y, 1) = \frac{n}{k} \{1 - G_2(Y_{n, n - \lfloor k y \rfloor})\}. 
\] 

Similarly for the bootstrap process \(\mathbb{F}_n^b\), 
we have 
$
\Phi({\tilde{R}^{b\prime}}/{q}) - \Phi({\tilde{R}^\prime}/{q}) = \hat{R}^{b\prime} - \hat{R}^\prime
$ on $ \mathbb{D}_{T}$, 
where \(\hat{R}^{b\prime}(x, y, z)\) is the bootstrap estimator for the quasi-tail copula, 
\[ 
\hat{R}^{b\prime}(x, y, z) := \frac{1}{k} \sum_{i=1}^{\lfloor nz \rfloor} \xi_{bi} \mathbf{1}\left\{X_i^{(n)} > U_b^{(1)}\left(\frac{n}{k x}\right), Y_i^{(n)} > U_b^{(2)}\left(\frac{n}{k y}\right) \right\}. 
\]


\subsection{Estimator for the Integrated Scedasis Functions.}
We propose the estimator for the integrated scedasis functions $C_j$ for $z\in[0,1]$ by
\begin{equation}\label{est:c}
 \hat{C}_{1} (z) := \frac{1}{k_1}\sum_{i=1}^{\lfloor nz \rfloor} \mathbf{1}\left(X_i^{(n)} > X_{n- k_1,n}\right),\quad 
 \hat{C}_{2} (z) := \frac{1}{k_2}\sum_{i=1}^{\lfloor nz \rfloor} \mathbf{1}\left(Y_i^{(n)} > Y_{n- k_2,n}\right).
\end{equation}
 Notice that the intermediate orders for the two margins are different. 
 By plugging $(k_1 / k, \infty, z)$ into the process $\hat{R}^\prime$,
 we have that 
\[
 \hat{R}^\prime (k_1 / k, \infty, z) = \frac{1}{k} \sum_{i=1}^{\lfloor nz \rfloor} \mathbf{1}\left(X_i^{(n)} > X_{n- k_1,n}\right)- \frac{k_1}{k} C_1(z) .
\]
Similarly, by plugging $(k_1 / k, \infty, z)$ into the process $\hat{R}^{b \prime}$, we derive 
$
\sum_{i=1}^{\lfloor nz \rfloor}\xi_{bi}\mathbf{1}(X_i^{(n)} > U_b^{(1)}({n}/{k_1}))/k_1$, and $
\sum_{i=1}^{\lfloor nz \rfloor}\xi_{bi}\mathbf{1}(Y_i^{(n)} > U_b^{(2)}({n}/{k_2}))/k_2 .
$
Notice the two processes do not equal $1$ when $z=1$, so we modify and derive the following bootstrap estimators.
\[
 \hat{C}^{b}_1(z) :=\frac{\sum_{i=1}^{\lfloor nz \rfloor}\xi_{bi}\mathbf{1}\left(X_i^{(n)} > U_b^{(1)}({n}/{k_1})\right)}
 {\sum_{i=1}^{n}\xi_{bi}\mathbf{1}\left(X_i^{(n)} > U_b^{(1)}({n}/{k_1})\right)}
 , \quad
 \hat{C}^{b}_2(z) :=\frac{\sum_{i=1}^{\lfloor nz \rfloor}\xi_{bi}\mathbf{1}\left(Y_i^{(n)} > U_b^{(2)}({n}/{k_2})\right)}{
    \sum_{i=1}^{n}\xi_{bi}\mathbf{1}\left(Y_i^{(n)} > U_b^{(2)}({n}/{k_2})\right)
 } .
\]

\begin{rem}
 \cite{Zhou2016} provide the estimators $\hat{c}_j(z)$ for $c_j(z)$. Based on their research, we can define a local estimator for $R(x,y,z)$ by the kernel methods. However, since this is another independent research direction and the primary focus of this study is on the bootstrap-based inference derived from the B-STEP, we leave the analysis of local estimators as a topic for future study.

 \end{rem}

\subsection{Estimators for the Extreme Value Indices.}
The well-known Hill estimators for $\gamma_j$ are given by
\begin{equation}
 \hat{\gamma}_1 = \frac{1}{k_1} \sum_{i=0}^{k_1} \log(X_{n-i,n}) - \log(X_{n-k_1,n}), \quad 
 \hat{\gamma}_2 = \frac{1}{k_2} \sum_{i=0}^{k_2} \log(Y_{n-i,n}) - \log(Y_{n-k_2,n}).
\end{equation}
To study the joint asymptotic properties of $\hat\gamma_j$ with other estimators and to propose the bootstrap estimators, it is necessary to rewrite $\hat\gamma_j$ as a functional of the tail empirical process. We start by taking $( n(1-G_1(U_1(n/k_1) x^{-\gamma_1})) / k, \infty, 1)$ into the process \eqref{eq: step-general} and derive the following tail empirical process
\begin{equation*}
 \mathbf{F}_{1n}(x) :=\frac{1}{q_1(x) k_1} \sum_{i=1}^{n} 
 \mathbf{1}{\left(X_i^{(n)}> x^{-\gamma_1} U_1(n / k_1)\right)} .
\end{equation*}
We then define the functional for a non-decreasing function $\theta$ on $\mathbb{R}$ with $\theta(0+) = 0$,
\begin{equation}\label{eq: psi}
  \Psi(\theta) := \int_{0}^{ ( q_1 \cdot \theta)^\leftarrow(1)} \theta(t) q_1(t) \frac{dt}{t}.
\end{equation}
Notice that for $\Pi(x) = {x}/{q_1(x)}$, it holds that $\Psi(\Pi) = 1$. Thus, the Hill estimators can be written as 
\begin{align*}
 \hat{\gamma}_1 &= 
 \gamma_1 \int^{\left(X_{n-k_1, n} / U_1(n / k_1)\right)^{-1/\gamma_1}}_{0} 
 \frac{1}{k_1} \sum_{i=1}^{n} 
 \mathbf{1}{\left(X_i^{(n)}> s^{-\gamma_1} U_1(n / k_1)\right)} \frac{d s}{s} 
 =\gamma_1 \Psi\left(\mathbf{F}_{1n}\right).
\end{align*}
Based on this functional, the bootstrap Hill estimator for $\hat\gamma_1$ is given by
\begin{align*}
 \hat{\gamma}_{1}^{b} := & \gamma_1 \, \Psi\left( \mathbf{F}_{1n}^b \right) 
 = \, \frac{1}{k_1} \sum_{i=1}^{n} \xi_{bi}\left(\log(X_{i}^{(n)}) - \log(U_{b}^{(1)}(n/k_1))\right) \mathbf{1}\left(X_{i}^{(n)} > U_{b}^{(1)}(n/k_1)\right) .
\end{align*}
with the bootstrap tail empirical process
\[
 \mathbf{F}_{1n}^b(x) := \frac{1}{q_1(x) k_1} \sum_{i=1}^{n} \xi_{bi} \mathbf{1}\left\{X_i^{(n)} > x^{-\gamma_1} U_1(n / k_1 ) \right\},
\]
Similarly, the bootstrap estimator for $\hat{\gamma}_2$ is given by 
\[ \hat{\gamma}_{2}^{b} := \frac{1}{k_2} \sum_{i=1}^{n} \xi_{bi}\left(\log(Y_{i}^{(n)}) - \log(U_{b}^{(2)}(n/k_2))\right) \mathbf{1}\left(Y_{i}^{(n)} > U_{b}^{(2)}(n/k_2)\right) .\]

We conclude this section by providing the derivatives of $\Phi$ and $\Psi$. These derivatives are crucial in the proof of the weak convergence of the three estimators. Generally, if a statistic is a functional of the B-STEP and the functional is Hadamard differentiable, the asymptotic properties of the statistic can be derived from the weak convergence of the B-STEP using the functional delta method. The concept of Hadamard differentiability is introduced in Chapter 3.9 of \cite{Van1996}. 
To proceed, we denote the following two functional classes:
\begin{align*}
 \mathcal{C}_R &:= \left\{\theta \in \ell^\infty\left(\mathbb{D}_{T}\right) \mid \theta \text { continuous with } \theta(\cdot, 0,\cdot)=\theta(0, \cdot, \cdot)=0\right\}, \\
 \mathcal{C}_{H,T} &:= \left\{\theta \in \ell^\infty\left([0,T]\right) \mid \theta \text { continuous with } \theta(0) = 0 \right\} .  
\end{align*}

\begin{prop}\label{prop:deri1}
Under Assumptions \ref{ass:marginal} and \ref{ass:cop},
$\Phi$ is Hadamard differentiable at $R^\prime / q$ tangentially to $\mathcal{C}_R$, 
whose derivative is the Lipschitz continuous functional,
\begin{align*}
  \Phi_{R^\prime / q}^{\prime}(\theta)(x,y,z)
=& \, q(x,y)\theta(x,y,z) \\
& \, -R^\prime_1(x,y,z) q_1(x)\theta\left(x, \infty, 1\right) - R^\prime_2(x,y,z) q_2(y) \theta\left(\infty, y, 1\right).
\end{align*}
\end{prop}

\begin{prop} \label{prop:deri2}
$\Psi$ is Hadamard-differentiable at \( \Pi(x)= x/q_1(x) \) tangentially to $\mathcal{C}_{H, 3/2}$ 
whose derivative is the Lipschitz continuous functional,
\begin{align*}
  \Psi^\prime_{{\Pi}}(\theta)(x) = \int_{0}^{1} \theta(t)  q_1(t) \frac{dt}{t} - \theta(1).
\end{align*}  
\end{prop}

Proofs for Proposition \ref{prop:deri1} and \ref{prop:deri2} are referred to in the supplementary material.


\section{Asymptotic Theorems}\label{sec:inference}

In this section, we present the asymptotic theorem for the B-STEP and the three estimators introduced in the previous section. For clarity, we denote in this section 
\(
(\xi_{1}, \ldots, \xi_{n}) = (\xi_{b1}, \ldots, \xi_{bn}).
\)
We use the notation \( W_n \leadsto W \) in \( \ell^\infty(\mathbb{D}_{T}) \) to indicate the weak convergence of the process \( W_n \) to a tight process \( W \) in the metric space \( \ell^\infty(\mathbb{D}_{T}) \) as $n\to\infty$. For the bootstrap process, we establish the conditional weak convergence, as defined by \citet{kosorok2003}. 
For an asymptotically measurable process in \( \ell^\infty(\mathbb{D}_{T}) \)
\[
W_n := W_n\left(X^{(n)}_{1}, \ldots, X^{(n)}_{n}, Y^{(n)}_{1}, \ldots, Y^{(n)}_{n}, \xi_{1}, \ldots, \xi_{n} \right),
\] 
the conditional weak convergence of \( W_n \) to \(W\) is defined as 
\begin{equation}\label{eq:weakdef}
 \sup _{h \in \mathrm{BL}_1(\ell^\infty(\mathbb{D}_{T}))}\left|\Em_{\xi} h\left(W_n\right)-\Em h(W)\right| = o_\Pm(1),\quad\text{as } n\to\infty,
\end{equation} 
where \( \Em_{\xi} \) represents the conditional expectation 
given \( (X^{(n)}_{1}, \ldots, X^{(n)}_{n}, Y^{(n)}_{1}, \ldots, Y^{(n)}_{n}) \), and 
the function class \( \mathrm{BL}_1(\ell^\infty(\mathbb{D}_{T})) \) is given by 
\[
\mathrm{BL}_1(\ell^\infty(\mathbb{D}_{T}))=\left\{ h: \ell^\infty(\mathbb{D}_{T}) \rightarrow \mathbb{R} \mid \| h \|_{\infty} \leq 1, |h(f_1)-h(f_2)| \leq \| f_1 - f_2\|, \, \forall f_1 , f_2 \in \ell^\infty(\mathbb{D}_{T}) \right\}.
\] 
We denote the conditional weak convergence as 
$
W_n \underset{\xi}{\overset{\mathbb{P}}{\leadsto}} W$ in $\ell^\infty(\mathbb{D}_{T}).
$
For a comprehensive review of conditional weak convergence, we refer to \citet{Bucher2019} 
and \citet{Van1996}. Theoretically, equation \eqref{eq:weakdef} states that the expectation of the unknown distribution \( h(W) \) can be approximated by computing the conditional expectation \( h(W_n) \).

In the following, we assume the process \( W \) is a Weiner process on \( \mathbb{D}_{T} \) with covariance function
\begin{equation}\label{eq:process}
 \operatorname{cov} \left(W\left(x_1, y_1, z_1\right), W\left(x_2, y_2, z_2\right)\right) 
 = {R}^\prime\left( x_1 \wedge x_2 , y_1 \wedge y_2, z_1 \wedge z_2\right),
\end{equation}
where \( \infty \wedge \infty := \infty \). 
We now establish the asymptotic property of B-STEP.

\begin{thm} \label{thm: step}
 Under Assumptions \ref{ass:marginal}-\ref{ass:bootweight}, 
 we have that as \( n \to \infty \),
  \begin{equation}\label{eq:thm: step1}
 \sqrt{k} \mathbb{F}_n \leadsto W / q
 \quad \text{and} \quad 
 \sqrt{k} \mathbb{F}_n^b \underset{\xi}{\overset{\mathbb{P}}{\leadsto}} W / q \quad \text{in } \ell^\infty(\mathbb{D}_{T}).
   \end{equation}
\end{thm}

To prove Theorem~\ref{thm: step}, we first establish the weak convergence of the \textit{simple bivariate sequential tail empirical process} in Propositions~\ref{thm:ba} and~\ref{prop:Bootsimple}. 
Then, under Assumptions~\ref{ass:marginal}-\ref{ass:bootweight}, we prove that the $\ell^\infty$ distance between the B-STEP and the simple B-STEP converges in probability to $0$, thereby establishing the weak convergence of the B-STEP. 
The definition of the simple B-STEP, Propositions~\ref{thm:ba} and~\ref{prop:Bootsimple}, and the detailed proof of Theorem~\ref{thm: step} are provided in Section~\ref{sec: pfstep} of the supplementary material.

\begin{rem}
 For the conditional weak convergence, it is worth noting that Assumption \ref{ass: smoothing} is not necessary for the proof.
 Recall the relation 
  \[
 \sqrt{k} \mathbb{F}_n(x,y,z) = \frac{\sqrt{k}}{q(x,y)} \left|
 \tilde{R}^\prime(x,y,z) - \Em(\tilde{R}^\prime(x,y,z)) +
 \Em (\tilde{R}^\prime(x,y,z)) - R^\prime(x,y,z) 
 \right|.
  \]
 In the proof, we bound the asymptotic bias $\sqrt{k} \Em(\tilde{R}^\prime(x,y,z)) - R^\prime(x,y,z)$ by
  \[
 \sup_{(x,y,z) \in \mathbb{D}_{T}}
 \frac{\sqrt{k} }{q(x,y)} \left| \sum_{i=1}^{ \lfloor nz\rfloor } \frac{1}{n}\frac{n}{k} C_{n,i}\left(\frac{kc_1(i/n)x}{n},\frac{kc_2(i/n)y}{n} \right) - R^\prime(x,y,z) \right| .
  \]
 Assumption \ref{ass: smoothing} is utilized to demonstrate that the above term converges to $0$ as $n\to\infty$. Since the bootstrap estimator is based on the process
  \[\sqrt{k} \mathbb{F}^{b}_n(x, y, z) =  \frac{ 1}{\sqrt{k} q(x,y)} \sum_{i=1}^{\lfloor nz \rfloor} \left({\xi_{bi}} - 1\right)\mathbf{1}\left\{X_i^{(n)} > U_1(n / k x), Y_i^{(n)} > U_2(n / k y) \right\}\]
the bias item vanishes in the proof.
\end{rem}

We then establish the asymptotic results for the integrated scedasis function and the quasi-tail copula. 
Define \( W_R \) by
\[
 W_R(x,y,z) := W(x,y,z)- R_1^\prime(x,y,z) W\left(x, \infty, 1\right)- R^\prime_2(x,y,z) W\left(\infty, y, 1\right),
\]
\(W_C^{(1)}(z) = s_1 W_R(s_1^{-1},\infty,z) \), and \(W_C^{(2)}(z) = s_2 W_R(\infty, s_2^{-1},z) \).

\begin{thm} \label{cor: quasi-tail}
 Under Assumptions \ref{ass:marginal}–\ref{ass:bootweight}, we have that for \( j = 1,2 \) and as \( n \to \infty \),
  \begin{equation}
 \sqrt{k} \left(\hat{R}^\prime - R^\prime \right) \leadsto W_R , 
 \quad \text{and} \quad 
 \sqrt{k} \left(\hat{R}^{b\prime} - \hat{R}^\prime \right) \underset{\xi}{\overset{\mathbb{P}}{\leadsto}} W_R
 \quad \text{in } \ell^\infty(\mathbb{D}_{T}).
  \end{equation}
  \begin{equation}
\sqrt{k} \left(\hat{C}_j - C_j \right) \leadsto W_C^{(j)} 
 \quad \text{and} \quad 
 \sqrt{k} \left(\hat{C}_j^b - \hat{C}_j \right) \underset{\xi}{\overset{\mathbb{P}}{\leadsto}} W_C^{(j)}
 \quad \text{in } \ell^\infty([0,1]).
  \end{equation}
\end{thm}


Next, we prove the asymptotic result of the bootstrap Hill estimator. Denote \begin{align*}
  \Gamma_1 &:= s_1 \gamma_1 \left(\int_{0}^{1} W\left({t}{s^{-1}_1}, \infty, 1\right) \frac{dt}{t} - W\left({s_1^{-1}}, \infty, 1\right) \right), \\
  \Gamma_2 &:= s_2 \gamma_2 \left(\int_{0}^{1} W\left( \infty,{t}{s^{-1}_2}, 1\right) \frac{dt}{t} - W\left(\infty,{s_2^{-1}}, 1\right) \right).
\end{align*}

\begin{thm}\label{cor: hill}
 Under Assumptions \ref{ass:marginal}-\ref{ass:bootweight}, 
 we have that for $j=1,2$ and as $n \rightarrow \infty$, 
  \begin{equation}
 \sqrt{k}(\hat{\gamma}_j - {\gamma}_j) {\leadsto} \Gamma_j, \quad \text{and} \quad
 \sqrt{k}(\hat{\gamma}_j^b - \hat{\gamma}_j) 
 \underset{\xi}{\overset{\mathbb{P}}{\leadsto}} \Gamma_j
 \quad \text{in \,} \mathbb{R}.
  \end{equation}
\end{thm}


\begin{rem}
 The asymptotic covariance of $(\sqrt{k} (\hat{\gamma}_1 - \gamma_1), \sqrt{k} (\hat{\gamma}_2-\gamma_2))$ is given by
\[
  \begin{bmatrix}
 s_1 \gamma_1^2       & R^\prime(s_2,s_1,1) \gamma_1\gamma_2\\
 R^\prime(s_2,s_1, 1) \gamma_1\gamma_2 & s_2 \gamma_2^2    \end{bmatrix}.
\]
In applications, we estimate $R^\prime(s_2,s_1, 1)$ by $k^2\hat{R}^\prime(k_1/ k, k_2/ k,1) / (k_1 k_2)$. 
\end{rem}

\section{Bootstrap-Based Tests}\label{sec:Test}

In this section, we formally define three hypothesis testing problems and subsequently propose a bootstrap-based approach for conducting these tests. Following this, we design comprehensive simulation experiments to evaluate the performance of the proposed tests.For each realization of $\{(X_i^{(n)}, Y_i^{(n)})\}_{i =1}^{n}$, we simulate $\xi_b$ and $\hat{\gamma}_j^{b}$, $\hat{C}_j^{b}$, $\hat{R}^{b\prime}$ for $b=1,2,\ldots, B$ as defined in Section~\ref{sec:tailst}. We denote the significance level as $\alpha$ in this section.


\subsection{Test for Equal Tail Heaviness.}
The first test examines whether $\{X_i^{(n)}\}$ and $\{Y_i^{(n)}\}$ exhibit the same tail heaviness without assuming a prior knowledge of the changing dependence structure $C_{n, i}$ and the scedasis functions $c_1$ and $c_2$. This hypothesis test is
\begin{equation}\label{eq:null}
 H_{10}:\gamma_1=\gamma_2 \quad \text{vs.} \quad H_{11}: \gamma_1\neq\gamma_2.
\end{equation}


To formulate the bootstrap-based test, we define 
\begin{align*}
 T_{H10} = k \left(\hat{\gamma}_1 - \hat{\gamma}_2\right)^2, \quad T_{H10}^{b} = k \left(\hat{\gamma}^b_1 - \hat{\gamma}^b_2 - \hat{\gamma}_1 + \hat{\gamma}_2\right)^2.
\end{align*}
The empirical distribution of the bootstrap samples is defined as \( F_{H10}(x) := \frac{1}{B} \sum_{b=1}^{B} \mathbf{1}(T_{H10}^b \le x) \), and its inverse function is defined as 
\( \hat{u}_{10}(\alpha) = F_{H10}^{\leftarrow}(\alpha) \).
We reject the null hypothesis if $T_{H10} \ge \hat{u}_{10}(1-\alpha)$.

The bootstrap method can effectively address cases where the two samples exhibit asymptotically tail independence.
For instance, when $R(1,1,z) \equiv 0$, the asymptotic behavior of the bivariate sequential tail empirical process, as established in Theorem \ref{thm: step}, becomes less reliable.
However, given that the covariance between $W(x,\infty,z)$ and $W(\infty,y,z)$ is $0$ for $x,y \in [0,1]^2$, the two Hill estimators become asymptotically independent.
Consequently, the functional delta method and continuous mapping theorem remain applicable to the marginal statistics, thereby ensuring the validity of the asymptotic results presented in Theorem \ref{cor: hill}. This finding indicates the potential effectiveness and conciseness of the bootstrap method as an inference tool under heteroscedastic extremes.

\subsection{Test for Equal Scedasis Functions.}
The second hypothesis test evaluates whether the two marginal distributions have the same scedasis function, i.e.,
\begin{equation}\label{eq:null4}
 H_{20}:c_1=c_2, \quad \forall\, t \in [0,1] \quad \text{vs.} \quad H_{21}:c_1 \neq c_2, \quad \exists\, t \in [0,1].
\end{equation} 
In financial applications, this test helps to determine whether two assets undergo identical crises. Variations in the scedasis function, which capture the impact of financial crises, play a crucial role in this assessment \citep{Zhou2016}.
Specifically, we define the test statistics,
\[
T_{20}(z) = \hat{C}_1(z) - \hat{C}_2(z), \quad T^{b}_{20}(z) = \hat{C}^{b}_1(z) - \hat{C}^{b}_2(z).
\]
The KS and Cram\'er-von Mises (CVM) statistics are defined as
    \begin{align*}     
 T_{H20}^{\mathrm{(KS)}} &= \sup_{z \in [0,1]} \sqrt{k} \left| T_{20}(z) \right|,
& 
 T_{H20}^{b\mathrm{(KS)}} & = \sup_{z \in [0,1]} \sqrt{k} \left| T^{b}_{20}(z) - T_{20}(z)\right|, \\
 T_{H20}^{\mathrm{(CVM)}} &= {k} \int_{0}^{1} \left( T_{20}(z) \right)^2 \, dz, 
& T_{H20}^{b\mathrm{(CVM)}} &= {k} \int_{0}^{1}\left( T^{b}_{20}(z) - T_{20}(z) \right)^2 \, dz,
    \end{align*}
 respectively, with the bootstrap distributions given by
    $$
 F_{H20}^{\mathrm{(KS)}}(x) := \frac{1}{B} \sum_{b=1}^{B} \mathbf{1}(T_{H20}^{b\mathrm{(KS)}} \le x), \quad
 F_{H20}^{\mathrm{(CVM)}}(x) := \frac{1}{B} \sum_{b=1}^{B} \mathbf{1}(T_{H20}^{b\mathrm{(CVM)}} \le x).
    $$
 The corresponding quantile function is denoted as
    $
 \hat{u}_{20}^{\mathrm{(KS)}}$, $\hat{u}_{20}^{\mathrm{(CVM)}}$. 
We reject the null hypothesis when $T_{H20}^{\mathrm{(KS)}} \ge \hat{u}_{20}^{\mathrm{(KS)}}(\alpha)$ and $T_{H20}^{\mathrm{(CVM)}} \ge \hat{u}_{20}^{\mathrm{(CVM)}}(\alpha)$ for KS and CVM tests, respectively.

Note that when \( C_1 = C_2 \) and the copula changes across sample, the process \( \hat{C}_1 - \hat{C}_2 \) does not converge to a Brownian bridge. For example, if we examine the covariance structure of \( W_C^{(1)} - W_C^{(2)} \) for \( 0 \leq z_1 \leq z_2 \leq 1 \), by assuming \( s_1 = s_2 = 1 \), we find that
\begin{align*}
& \operatorname{cov} 
\left(W_{C}^{(1)}(z_1) - W_{C}^{(2)}(z_1), W_{C}^{(1)}(z_2) - W_{C}^{(2)}(z_2)\right) \\
=& 2 C_1(z_1)(1 - C_1(z_2)) + 2\left[C_1(z_2)R^\prime(1,1,z_1) + C_1(z_1)R^\prime(1,1,z_2)\right] \\[6pt]
& - 2\left[R^\prime(1,1,z_1) + C_1(z_1)C_1(z_2)R^\prime(1,1,1)  \right].
\end{align*}
Consequently, the bootstrap method becomes essential for conducting the test.
A similar result is also observed in the CVM test for
$
 k \int_{0}^{1} \left(\hat{C}_1(z) - \hat{C}_2(z)\right)^2 dz.
$





\subsection{Test for a Non-Changing Tail Copula.}
The final test assesses whether the tail copula remains unchanged under identical scedasis functions:
  \begin{align}
   & H_{30}: R'(x,y,z) = R(x,y,1), \quad \forall (x,y,z) \in \mathbb{D}_T \nonumber \\
 \text{vs. } & 
 H_{31}: R'(x,y,z) \neq R(x,y,1), \quad \exists (x,y,z) \in \mathbb{D}_T. \label{eq:null5}
  \end{align}
This test serves as a validation of the model proposed in \cite{einmahl2024tail}. Furthermore, compared to the test described in \cite{drees2023statistical}, our test represents a specific case of assessing non-changing copulas when \( C_1 = C_2 \). To formulate the test statistics, denote 
\[
T_{30}(z) = \hat{C}_1(z) + \hat{C}_2(z) - \frac{2\hat{R}^\prime(k_1 / k, k_2/ k,z)}{\hat{R}^\prime(k_1 / k,k_2 / k,1)}, \quad
T_{30}^b(z)  = \hat{C}^{b}_1(z) + \hat{C}^{b}_2(z) - \frac{2\hat{R}^{b\prime}(k_1/ k ,k_2 / k,z)}{\hat{R}^{b\prime}(k_1/ k ,k_2 / k,1)},
\]
and the KS and the CVM statistics are given by
\begin{align*}     
T_{H30}^{\mathrm{(KS)}} &= \sup_{z \in (0,1]} \sqrt{k} \left|T_{30}(z) \right|,&T_{H30}^{b\mathrm{(KS)}} &= \sup_{z \in (0,1]} \sqrt{k} \left| T_{30}^b(z) - T_{30}(z) \right|, \\
T_{H30}^{\mathrm{(CVM)}} &= {k} \int_{0}^{1} \left(T_{30}(z) \right)^2 \, dz, & 
T_{H30}^{b\mathrm{(CVM)}} &= k \int_{0}^{1} \left( T_{30}^b(z) - T_{30}(z) \right)^2 \, dz,
   \end{align*}
respectively, with the bootstrap distributions given by
   \[
F_{H30}^{\mathrm{(KS)}}(x) := \frac{1}{B} \sum_{b=1}^{B} \mathbf{1}\left(T_{H30}^{b\mathrm{(KS)}} \le x\right), \quad 
F_{H30}^{\mathrm{(CVM)}}(x) := \frac{1}{B} \sum_{b=1}^{B} \mathbf{1}\left(T_{H30}^{b\mathrm{(CVM)}} \le x\right),
   \]
and the corresponding quantiles as $
\hat{u}_{30}^{\mathrm{(KS)}}$ and $
\hat{u}_{30}^{\mathrm{(CVM)}}.$

The test statistics based on \( T_{30}(z) \) converge in distribution to their limits faster than the integrated angular measures \citep[Corollary 3.2]{drees2023statistical}. Specifically, given \( C_1 = C_2 \) and \( H_{30} \), we have that as \( n \to \infty \),
\begin{align*}
 \sup_{z \in [0,1]} \sqrt{k} 
 \left| T_{30}(z) \right| 
  \leadsto 
 \sup_{z \in [0,1]} \left| \left\{4(R(s_1^{-1},s_2^{-1},1))^{-1} - 3 s_1 - 3 s_2 + 2 R(s_2, s_1,1)\right\} B(z) \right|,
\end{align*}
where \( B \) is a standard Brownian bridge. Consequently, our method requires less data to test \eqref{eq:null5} compared to \cite{drees2023statistical} in the special case \( C_1 = C_2 \).

\begin{prop}\label{thm:hilltestboot}
 Under Assumptions \ref{ass:marginal}-\ref{ass:bootweight}, for $\mathrm{M} \in \{\mathrm{CVM},\mathrm{KS}\}$, as \( n \to \infty \) and \( B \to \infty \), we have that
  \begin{enumerate}[\rm (a) ]
    \item if \( H_{10} \) holds, then 
    $
\Pm(T_{H10} \ge \hat{u}_{10}(1-\alpha)) \to \alpha;
    $
    \item if \( H_{20} \) holds, then
    $
\Pm(T_{H20}^{\mathrm{(M)}} \ge \hat{u}_{20}^{\mathrm{(M)}}(1-\alpha)) \to \alpha$;
    \item if \( c_1 = c_2 \) and \( H_{30} \) holds,
    $
\Pm(T_{H30}^{\mathrm{(M)}} \ge \hat{u}_{30}^{\mathrm{(M)}}(1-\alpha)) \to \alpha.
$
  \end{enumerate}
\end{prop}


\subsection{Simulation Results}

We now turn to the finite sample performance of the three proposed tests. The simulation study comprises 1000 replications for each setting, with each replicate undergoing 200 bootstrap iterations. We set the sample size $n = 1000$ and the intermediate order $k = 50, 100, 200$. All tests are conducted at $\alpha = 0.05$ and $0.01$.
The random weights $\xi_{bi}$ are generated from a standard exponential distribution.

The two marginal scedasis functions for $ z \in [0, 1]$ are chosen from the following options: 
\[\begin{array}{lll}
 a_1(z) = 0.8 + 0.4 z, \quad & a_2(z) = 0.6 + 0.8 z, \\ 
 a_3(z) = 1 + 0.5 \sin(2\pi z), \quad & a_4(z) = 1 + 0.2 \sin(2\pi z), \\ 
 a_5(z) = 
 \begin{cases} 
 0.5 + 2z, & z \in \left[0, \frac{1}{2}\right] \\ 
 2.5 - 2z, & z \in \left(\frac{1}{2}, 1\right] 
 \end{cases}, 
 & a_6(z) = 
 \begin{cases} 
 0.25 + 3z, & z \in \left[0, \frac{1}{2}\right] \\ 
 3.25 - 3z, & z \in \left(\frac{1}{2}, 1\right] 
 \end{cases}. &
\end{array}\]
These functions have been utilized in prior research, such as \cite{einmahl2024tail} and \cite{drees2023statistical}. Among them, $a_1$ and $a_2$ are linear functions, $a_3$ and $a_4$ follow sinusoidal patterns, while $a_5$ and $a_6$ are piecewise linear functions. For $j = 1,2,3$, the even-indexed functions $a_{2j}$ display volatility characteristics akin to their odd-indexed counterparts $a_{2j-1}$. However, the fluctuation behavior of $a_1$, $a_3$, and $a_5$ differs markedly. In our experiment, we examine how these similarities and differences in scedasis functions impact the two-sample test.


In all cases, the marginal distributions follow a Fr\'echet distribution such that 
\[
F_{n,i}^{(j)}(x) = \exp(-c_j(i/n) x^{-\lambda_j}),
\] 
where $c_j$ is the scedasis function and $\lambda_j$ is a shape parameter. The Fr\'echet distribution with shape parameter $\lambda_j$ has an EVI of $1/\lambda_j$. 
Since the test statistics of $H_{20}$ and $H_{30}$ are rank-based, the marginal distributions do not influence their results. Therefore, for these tests, we fix $\lambda_j = 2$. For the test assessing the equivalence of EVIs, the values of $\lambda_j$ are provided in Table \ref{tab:hilltest}. 


On the other hand, we construct the changing copula by combining the $ t$ copula and the GumbelcCopula. 
Denote the bivariate $t$-copula as $\mathcal{C}_{t}(u, v; \nu, \rho)$ with degrees of freedom $\nu$ and correlation $\rho$, and 
the Gumbel copula with a dependence parameter $\theta\geq 1$ as $\mathcal{C}_{g}(u, v; \theta)$, which is given by 
\[
\mathcal{C}_{g}(u, v; \theta) = \exp\left[-\left\{(-\ln u)^{\theta} + (-\ln v)^{\theta}\right\}^{1/\theta}\right], \quad u, v \in \mathbb{R}_+,
\] 
Based on these two copula families, we define six changing (survival) copula for our tests.
For $j = 1,2,5$, we define 
\[
\mathcal{C}_j(x, y, z) = p_j(z) \cdot \mathcal{C}_t(x, y; 2, 0) + (1 - p_j(z)) \cdot \mathcal{C}_g(x, y; 2), \quad x,y \in \mathbb{R}_+, z \in [0,1],
\] 
where $p_1(z) = 0.5 + 0.5 \cos(2\pi z)$, $p_2(z) = 0.5 + 0.2 \cos(2\pi z)$, $p_5 \equiv 0.5$ for $z \in [0,1]$.
These copulas possess the characteristic that the mixture weight $p_j(t)$ changes smoothly across $z$.
For $j = 3,4,6$, we define 
\[
\mathcal{C}_j(x, y, z) = \mathcal{C}_g(x, y; \theta = p_j(z)), \quad x, y \in \mathbb{R}_+,\, z \in [0,1],
\] 
where $p_3(z) = 2 + 3z$, $p_4(z) = 2 + z$, and $p_6 \equiv 2$ for $ z \in [0,1]$.
These copulas possess the characteristic that the dependence parameter $\theta$ of the Gumbel copula varies as a function of $z$. 

\begin{table}[htbp]
  \centering
  \caption{Simulated rejection frequency for testing $H_{10}$.}
   \begin{tabular}{rrlllrrrrrrr}
   \toprule
   \multicolumn{2}{c}{EVIs} & \multicolumn{2}{c}{scedasis} &  & \multicolumn{1}{l}{$k$} & \multicolumn{2}{l}{50} & \multicolumn{2}{l}{100} & \multicolumn{2}{l}{200} \\
 \cmidrule{6-12}  \multicolumn{1}{l}{$\lambda_1$} & \multicolumn{1}{l}{$\lambda_2$} & $c_1$ & $c_2$ & & \multicolumn{1}{l}{$\alpha$} & \multicolumn{1}{l}{0.05} & \multicolumn{1}{l}{0.1} & \multicolumn{1}{l}{0.05} & \multicolumn{1}{l}{0.1} & \multicolumn{1}{l}{0.05} & \multicolumn{1}{l}{0.1} \\
   \midrule
 2.5 & 2.5 & $a_1$ & $a_3$ & $\mathcal{C}_1$ &  & 0.03 & 0.07 & 0.04 & 0.08 & 0.05 & 0.10 \\
 2 & 2 & $a_1$ & $a_5$ & $\mathcal{C}_2$ &  & 0.03 & 0.07 & 0.04 & 0.08 & 0.05 & 0.10 \\
 3 & 3 & $a_3$ & $a_5$ & $\mathcal{C}_3$ &  & 0.02 & 0.06 & 0.03 & 0.07 & 0.03 & 0.06 \\
 2.5 & 2 & $a_1$ & $a_3$ & $\mathcal{C}_4$ &  & 0.34 & 0.47 & 0.69 & 0.79 & 0.95 & 1.00 \\
 2.5 & 2.2 & $a_1$ & $a_5$ & $\mathcal{C}_1$ &  & 0.11 & 0.19 & 0.20 & 0.31 & 0.38 & 0.53 \\
 2.5 & 2.4 & $a_3$ & $a_5$ & $\mathcal{C}_2$ &  & 0.04 & 0.10 & 0.06 & 0.12 & 0.07 & 0.13 \\
 2.5 & 2.6 & $a_1$ & $a_3$ & $\mathcal{C}_3$ &  & 0.04 & 0.08 & 0.05 & 0.09 & 0.07 & 0.12 \\
 2.5 & 2.8 & $a_1$ & $a_5$ & $\mathcal{C}_4$ &  & 0.10 & 0.19 & 0.22 & 0.33 & 0.41 & 0.53 \\
 2.5 & 3 & $a_3$ & $a_5$ & $\mathcal{C}_1$ &  & 0.17 & 0.28 & 0.34 & 0.47 & 0.64 & 0.76 \\
   \bottomrule
   \end{tabular}%
  \label{tab:hilltest}%
 \end{table}%

Table~\ref{tab:hilltest} provides a summary of the simulated rejection frequencies for the test \eqref{eq:null}. The first two columns specify the $\lambda_j$; the third and fourth columns represent the scedasis functions selected from ${a_1, a_3, a_5}$; the fifth column identifies the changing copula structures.
 For cases with equivalent EVIs (rows 1-4), the rejection frequencies at $k = 200$ closely match the significance levels (0.05 and 0.10), demonstrating that the test effectively controls Type I errors. However, when $k = 50$ or $k = 100$, the rejection frequencies are lower than the significance levels, suggesting that a larger $k$ is necessary to achieve accurate performance.
 For cases with differing EVIs (rows 5-9), the rejection frequencies increase with larger $k$. For example, in row 4 (where $\lambda_1 = 2.5$ and $\lambda_2 = 2$), the rejection frequency reaches 95\% at $\alpha = 0.05$ when $k = 200$, compared to only 34\% at $k = 50$. 
 
\begin{table}[htbp]
  \centering
  \caption{Simulated rejection frequency for testing $H_{20}$.}
 \small
 \begin{tabular}{lllrrrrrrrrrrrrr}
  \toprule
  \multicolumn{3}{c}{Distribution} & \multicolumn{1}{l}{$k$} & \multicolumn{4}{l}{50} & \multicolumn{4}{l}{100} & \multicolumn{4}{l}{200} \\
\cmidrule{4-16}  \multicolumn{2}{c}{scedasis} &  & \multicolumn{1}{l}{$\alpha$} & \multicolumn{2}{l}{0.05} & \multicolumn{2}{l}{0.1} & \multicolumn{2}{l}{0.05} & \multicolumn{2}{l}{0.1} & \multicolumn{2}{l}{0.05} & \multicolumn{2}{l}{0.1} \\
\cmidrule{4-16}  $c_1$ & $c_2$ &  & \multicolumn{1}{l}{M} & \multicolumn{1}{l}{KS} & \multicolumn{1}{l}{CVM} & \multicolumn{1}{l}{KS} & \multicolumn{1}{l}{CVM} & \multicolumn{1}{l}{KS} & \multicolumn{1}{l}{CVM} & \multicolumn{1}{l}{KS} & \multicolumn{1}{l}{CVM} & \multicolumn{1}{l}{KS} & \multicolumn{1}{l}{CVM} & \multicolumn{1}{l}{KS} & \multicolumn{1}{l}{CVM} \\
  \midrule
  $a_1$ & $a_1$ & $\mathcal{C}_1$ &  & 0.03 & 0.04 & 0.06 & 0.07 & 0.04 & 0.03 & 0.07 & 0.07 & 0.05 & 0.05 & 0.09 & 0.10 \\
  $a_2$ & $a_2$ & $\mathcal{C}_2$ &  & 0.03 & 0.04 & 0.07 & 0.07 & 0.03 & 0.04 & 0.08 & 0.09 & 0.05 & 0.05 & 0.10 & 0.10 \\
  $a_3$ & $a_3$ & $\mathcal{C}_3$ &  & 0.01 & 0.01 & 0.04 & 0.04 & 0.01 & 0.01 & 0.03 & 0.04 & 0.03 & 0.02 & 0.06 & 0.06 \\
  $a_4$ & $a_4$ & $\mathcal{C}_4$ &  & 0.01 & 0.02 & 0.03 & 0.03 & 0.01 & 0.02 & 0.03 & 0.04 & 0.03 & 0.03 & 0.06 & 0.06 \\
  $a_5$ & $a_5$ & $\mathcal{C}_5$ &  & 0.03 & 0.03 & 0.06 & 0.06 & 0.04 & 0.04 & 0.08 & 0.08 & 0.05 & 0.05 & 0.11 & 0.10 \\
  $a_6$ & $a_6$ & $\mathcal{C}_6$ &  & 0.01 & 0.02 & 0.04 & 0.04 & 0.02 & 0.03 & 0.05 & 0.06 & 0.03 & 0.03 & 0.07 & 0.06 \\
  $a_1$ & $a_2$ & $\mathcal{C}_1$ &  & 0.07 & 0.09 & 0.12 & 0.15 & 0.10 & 0.14 & 0.18 & 0.22 & 0.22 & 0.26 & 0.32 & 0.35 \\
  $a_3$ & $a_4$ & $\mathcal{C}_2$ &  & 0.13 & 0.16 & 0.22 & 0.24 & 0.30 & 0.32 & 0.44 & 0.42 & 0.60 & 0.58 & 0.70 & 0.70 \\
  $a_5$ & $a_6$ & $\mathcal{C}_3$ &  & 0.02 & 0.02 & 0.05 & 0.07 & 0.05 & 0.06 & 0.11 & 0.15 & 0.14 & 0.20 & 0.27 & 0.39 \\
  $a_1$ & $a_3$ & $\mathcal{C}_4$ &  & 0.86 & 0.90 & 0.94 & 0.95 & 1.00 & 1.00 & 1.00 & 1.00 & 1.00 & 1.00 & 1.00 & 1.00 \\
  $a_1$ & $a_5$ & $\mathcal{C}_5$ &  & 0.14 & 0.13 & 0.25 & 0.27 & 0.32 & 0.32 & 0.45 & 0.47 & 0.68 & 0.68 & 0.79 & 0.80 \\
  $a_3$ & $a_5$ & $\mathcal{C}_6$ &  & 0.64 & 0.68 & 0.76 & 0.79 & 0.96 & 0.97 & 0.98 & 0.99 & 1.00 & 1.00 & 1.00 & 1.00 \\
  \bottomrule
  \end{tabular}%
  \label{tab:H20}%
 \end{table}%

 Table~\ref{tab:H20} summarizes the simulated rejection frequencies for the test \eqref{eq:null4}. The first six rows correspond to scenarios under the null hypothesis, where we observe consistently low rejection frequencies.
 The subsequent six rows assess the power of the proposed methods under alternative hypotheses. When $k = 50, 100$, the rejection frequencies are relatively low for the seventh and ninth experiments. However, the rejection frequencies improve when $k = 200$, indicating that larger intermediate thresholds enhance the test's ability to detect differences in scedasis functions.
 Additionally, in most scenarios, the CVM test exhibits slightly higher rejection frequencies than the KS test.

\begin{table}[htbp]
  \centering
  \caption{Simulated rejection frequency for testing $H_{30}$.}
 \small
 \begin{tabular}{lllrrrrrrrrrrrrr}
  \toprule
  \multicolumn{3}{c}{Distribution } & \multicolumn{1}{l}{$k$} & \multicolumn{4}{l}{50} & \multicolumn{4}{l}{100} & \multicolumn{4}{l}{200} \\
\cmidrule{4-16}  \multicolumn{2}{c}{scedasis} &  & \multicolumn{1}{l}{$\alpha$} & \multicolumn{2}{l}{0.05} & \multicolumn{2}{l}{0.1} & \multicolumn{2}{l}{0.05} & \multicolumn{2}{l}{0.1} & \multicolumn{2}{l}{0.05} & \multicolumn{2}{l}{0.1} \\
\cmidrule{4-16}  $c_1$ & $c_2$ &  & \multicolumn{1}{l}{M} & \multicolumn{1}{l}{KS} & \multicolumn{1}{l}{CVM} & \multicolumn{1}{l}{KS} & \multicolumn{1}{l}{CVM} & \multicolumn{1}{l}{KS} & \multicolumn{1}{l}{CVM} & \multicolumn{1}{l}{KS} & \multicolumn{1}{l}{CVM} & \multicolumn{1}{l}{KS} & \multicolumn{1}{l}{CVM} & \multicolumn{1}{l}{KS} & \multicolumn{1}{l}{CVM} \\
  \midrule
  $a_1$ & $a_1$ & $\mathcal{C}_5$ &  & 0.03 & 0.02 & 0.07 & 0.07 & 0.03 & 0.03 & 0.08 & 0.07 & 0.04 & 0.04 & 0.09 & 0.09 \\
  $a_2$ & $a_2$ & $\mathcal{C}_6$ &  & 0.02 & 0.02 & 0.05 & 0.05 & 0.02 & 0.03 & 0.07 & 0.07 & 0.05 & 0.05 & 0.09 & 0.10 \\
  $a_3$ & $a_3$ & $\mathcal{C}_5$ &  & 0.03 & 0.04 & 0.08 & 0.08 & 0.03 & 0.03 & 0.07 & 0.07 & 0.08 & 0.08 & 0.13 & 0.16 \\
  $a_4$ & $a_4$ & $\mathcal{C}_6$ &  & 0.02 & 0.02 & 0.04 & 0.05 & 0.03 & 0.04 & 0.05 & 0.07 & 0.04 & 0.04 & 0.07 & 0.08 \\
  $a_1$ & $a_1$ & $\mathcal{C}_1$ &  & 0.12 & 0.15 & 0.24 & 0.27 & 0.29 & 0.36 & 0.47 & 0.55 & 0.62 & 0.69 & 0.78 & 0.85 \\
  $a_2$ & $a_2$ & $\mathcal{C}_3$ &  & 0.06 & 0.09 & 0.13 & 0.19 & 0.22 & 0.32 & 0.37 & 0.47 & 0.58 & 0.69 & 0.71 & 0.81 \\
  $a_3$ & $a_3$ & $\mathcal{C}_1$ &  & 0.12 & 0.14 & 0.23 & 0.29 & 0.31 & 0.39 & 0.48 & 0.58 & 0.62 & 0.75 & 0.80 & 0.89 \\
  $a_4$ & $a_4$ & $\mathcal{C}_3$ &  & 0.06 & 0.10 & 0.15 & 0.20 & 0.22 & 0.32 & 0.37 & 0.46 & 0.46 & 0.56 & 0.60 & 0.70 \\
  $a_1$ & $a_1$ & $\mathcal{C}_2$ &  & 0.04 & 0.04 & 0.08 & 0.09 & 0.05 & 0.06 & 0.12 & 0.13 & 0.08 & 0.07 & 0.19 & 0.16 \\
  $a_2$ & $a_2$ & $\mathcal{C}_4$ &  & 0.03 & 0.05 & 0.08 & 0.10 & 0.09 & 0.13 & 0.17 & 0.20 & 0.28 & 0.33 & 0.40 & 0.46 \\
  $a_3$ & $a_3$ & $\mathcal{C}_2$ &  & 0.04 & 0.05 & 0.09 & 0.09 & 0.05 & 0.07 & 0.12 & 0.14 & 0.10 & 0.14 & 0.21 & 0.25 \\
  $a_4$ & $a_4$ & $\mathcal{C}_4$ &  & 0.03 & 0.05 & 0.07 & 0.10 & 0.08 & 0.11 & 0.15 & 0.19 & 0.13 & 0.15 & 0.22 & 0.25 \\
  \bottomrule
  \end{tabular}%
  \label{tab:H30}%
 \end{table}%



 For the test of non-changing tail copula structure \eqref{eq:null5}, the experimental results are summarized in Table~\ref{tab:H30}. In this setting, $\mathcal{C}_5$ and $\mathcal{C}_6$ represent cases where the tail copula structure remains unchanged. Experiments with odd-numbered rows correspond to tests conducted under mixture copula settings, while even-numbered rows involve changing parameterized copula settings. The scedasis structures of the odd-numbered experiments are designed to resemble those of the even-numbered experiments.
 When $k = 100$, the test power is relatively low across all cases. However, when $k = 200$, the results improve significantly: the Type I error frequencies align more closely with the significance level, and the rejection frequencies for alternative hypotheses exceed the significance level.
 

\section{Conclusion}
This paper proposes a copula-based model for independent but not identically distributed data with bivariate heteroscedastic extremes. The model allows both the copula structure and distributions marginal to vary across samples. We develop a comprehensive framework to derive bootstrap estimators via the B-STEP process, using the functional delta method. Simulation results validate the robustness of our bootstrap approach.
Other bootstrap techniques (\citet{haan2024, Jentsch2021}) are not covered in our framework, and their asymptotic validity under bivariate heteroscedastic extremes remains unexplored. Extending our framework to time-series settings with mild dependence (e.g., mixing conditions as in \citet{10.1214/18-EJS1415, zou2021multiple}) is a potential next step, but the asymptotic analysis under such conditions is left as an open problem due to the need for a separate theoretical framework.

\newpage
\bibliographystyle{abbrvnat}
\bibliography{Bibliography-MM-MC}
\end{document}